\documentstyle[graphicx]{aipproc}

\begin{document}
\title{Tumour Therapy with Particle Beams}

\author{Claus Grupen\footnote{e-mail: grupen@aleph.physik.uni-siegen.de}}
\address{Department of Physics, University of Siegen \\ Germany}

\maketitle

\begin{abstract}
Photons are exponentially attenuated in matter producing high doses close
to the 
surface. Therefore they are not well suited for the treatment of deep seated 
tumours. Charged particles, in contrast, exhibit a sharp increase of
ionisation
density close to the end of their range, the so-called Bragg-peak. The
depth of 
the Bragg-peak can be adjusted by varying the particle's energy. In
parallel with 
the large energy deposit the increase in biological effectiveness for cell 
killing at the end of the range provides an ideal scalpel for the surgeon
effectively
without touching the surface tissue. Consequently proton therapy  has
gained 
a lot of ground for treating well localized tumours. Even superior still
are heavy 
ions, where the ionisation pattern is increased by the square of their charge 
$(\sim z^2)$.
\end{abstract}

\section*{Introduction}
It has been known for a long time that tissue, in particular tumour tissue,
is 
sensitive to ionising radiation. Therefore it is only natural that tumours
have been treated 
with various types of radiation, like $\gamma$-rays and electrons.
$\gamma$-rays 
are easily available from radioactive sources, like $^{60}$Co, and
electrons can 
be accelerated to $MeV$-energies by relatively inexpensive linear
accelerators.
The disadvantage of $\gamma$-rays and electrons is that they deposit most of 
their energy close to the surface. To reduce the surface dose in tumour
treatment 
requires rotating the source or the patient so that the surface dose is 
distributed over a larger volume. In contrast, protons and heavy ions deposit 
most of their energy close to the end of their range (Bragg-peak). The
increase 
in energy loss at the Bragg-peak amounts to a factor of about 5 compared to
the 
surface dose, depending somewhat on the particle's energy. Heavy ions
offer, in 
addition, the possibility to monitor the destructive power of the beam by 
observing annihilation radiation by standard positron-emission tomography 
techniques (PET). The annihilation radiation is emitted by $\beta^+$-active 
nuclear fragments produced by the incident heavy ion beam itself. 

\section*{Energy loss of particles in tissue [1]}
A photon beam is attenuated in matter according to
\begin{equation}
  I(x) = I_0 e^{-\mu x} 
\end{equation}
where $I_0$ is the initial intensity and $I(x)$ the beam intensity at the
depth $x$. 
$\mu$ 
is the linear mass attenuation coefficient which depends on the photon
energy $E$ and 
the target  charge $Z$. $\mu (E)$ is shown in figure~1 for a target
composed of 
water, which is essentially equivalent to tissue. The main interaction
mechanisms 
which contribute to $\mu(E)$ are the photoelectric effect $(\sim
Z^5/E^{3.5})$, 
Compton scattering $(\sim (Z/E) \; ln\; E)$ and pair-production $(\sim Z^2 \; 
 ln\; E)$. For energies typical for radioactive sources $(\sim MeV)$ Compton 
 scattering dominates. The absorption profile of photons in matter exhibits a 
 peak close to the surface followed by an exponential decay. 
 
 Charged particles 
 suffer energy loss by ionisation. This energy loss is described by the
Bethe-Bloch 
 formula: 

\begin{figure}
\begin{center}
\includegraphics[height=0.35\textheight]{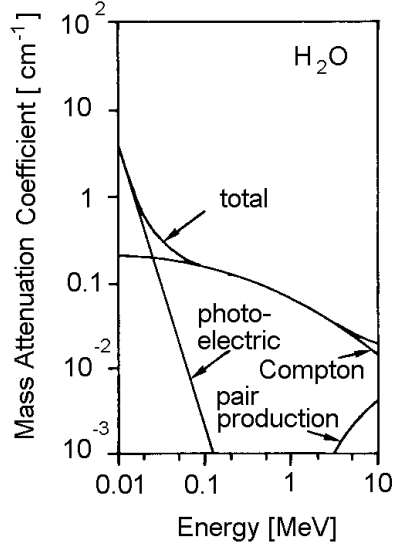}
\caption{Mass attenuation coefficient for photons in water as a function of
the 
photon energy \protect\cite{Grupen1}}
\end{center}
\end{figure}

\begin{equation}
  \frac{dE}{dx} = 2 \kappa \{ ln \frac{E^{{\rm max}}_{{\rm kin}}}{I} -
\beta^2 -
    \frac{\delta}{2} \}
\end{equation}
where
\begin{equation}
  \kappa = 2 \pi \, N\!\!_A \, r_e^2 m_e c^2 z^2 \frac{Z}{A} \cdot
\frac{1}{\beta^2} \; .
\end{equation} 
\begin{center}
\begin{tabular}{lcl}
  $z$ & -- &  charge of the beam particle \\
  $Z$ & -- & charge of the absorber material \\
  $A$ & -- & mass number of the absorber material \\
  $m_e$ & -- & electron mass \\
  $c$ & -- & velocity of light  \\
  $N_A$ & -- & Avogadro's number \\
  $r_e$ & -- & classical electron radius \\
  $\beta$ & -- & velocity of the particle divided by $c$ \\
  $E^{{\rm max}}_{{\rm kin}}$ & -- & maximum transferable energy \\
          &    & to an atomic electron \\
  $I$ & -- & mean excitation energy of the target material \\
  $\delta$ & -- & density parameter
\end{tabular}
\end{center}
For protons $(z=1)$ interacting in water (or tissue) equation (2) can be 
approximated by
\begin{equation}
  \frac{dE}{dx} = 0,16 \cdot \frac{1}{\beta^2} \; ln \frac{E^{{\rm
max}}_{{\rm 
  kin}} [eV]}{100}  \quad \left[ \frac{MeV}{cm} \right] 
\end{equation}
where
\begin{equation}
  E^{{\rm max}}_{{\rm kin}} \approx 2 m_e c^2 \beta^2 \gamma^2 \; ,
\end{equation}
which gives an energy loss of 4.2 $MeV/cm$ for 200 $MeV$ protons at the
surface 
and $\sim 20 \; MeV/cm$ close to the end of their range. For heavy ions the
energy 
loss is essentially scaled by $z^2$. When charged particles reach the end
of their 
range the energy loss first rises like $1/\beta^2$ but when they are very
slow 
they capture electrons from the target material and their effective charge 
decreases and hence their energy loss rapidly falls to zero.

A typical energy loss curve for ions as a function of their energy is
sketched in 
figure~2 \cite{Kraft1}. The energy loss of $^{12}$C ions as a function of
the depth 
in water is shown in figure~3 \cite{Kraft1,Kraft2}. The tail of the energy
loss 
beyond the Bragg-peak originates from fragmentation products of $^{12}$C
ions, 
which are faster than the $^{12}$C ions and have a somewhat longer range.

In the ionisation process a generally small fraction of the particle's
energy is 
transferred to the atomic electrons. In rare cases these electrons can get
a larger 
amount of energy. The $\delta$-electrons deviate from the main ionisation
trail
and produce a fuzzy-like track (figure~4, \cite{Kraft1}).

\begin{figure}
\begin{center}
\includegraphics[height=0.3\textheight]{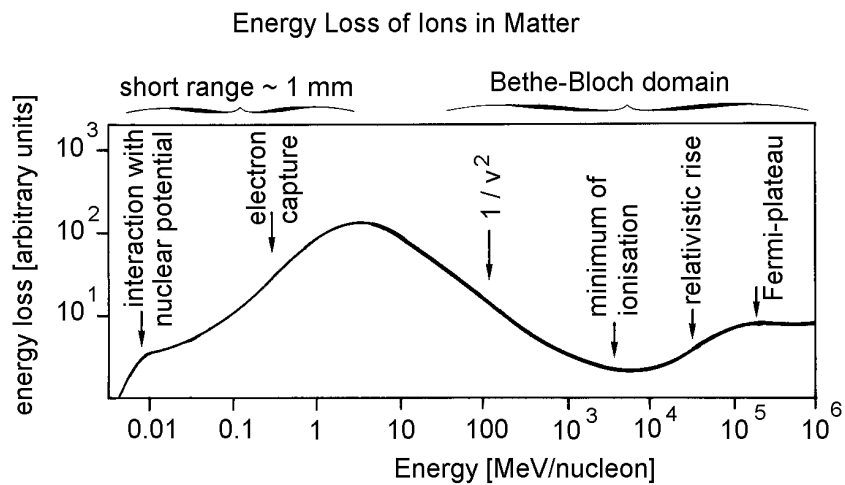}
\caption{Energy loss of ions in matter as a function of their energy (after 
\protect\cite{Kraft1})}
\end{center}
\end{figure}

\begin{figure}
\begin{center}
\includegraphics[height=0.275\textheight]{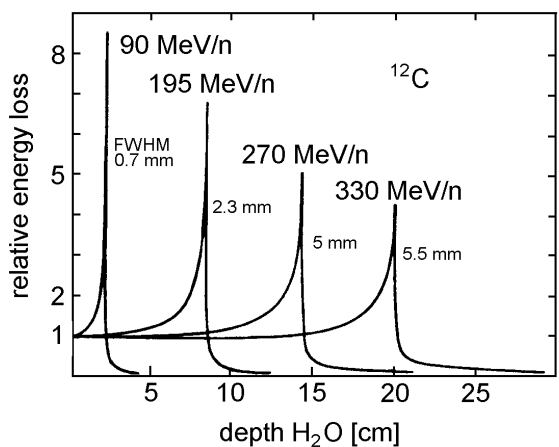}
\caption{Energy loss of carbon-ions ($^{12}$C) in water as a function of
depth
\protect\cite{Kraft1,Kraft2}}
\end{center}
\end{figure}

\begin{figure}
\begin{center}
\includegraphics[height=0.3\textheight]{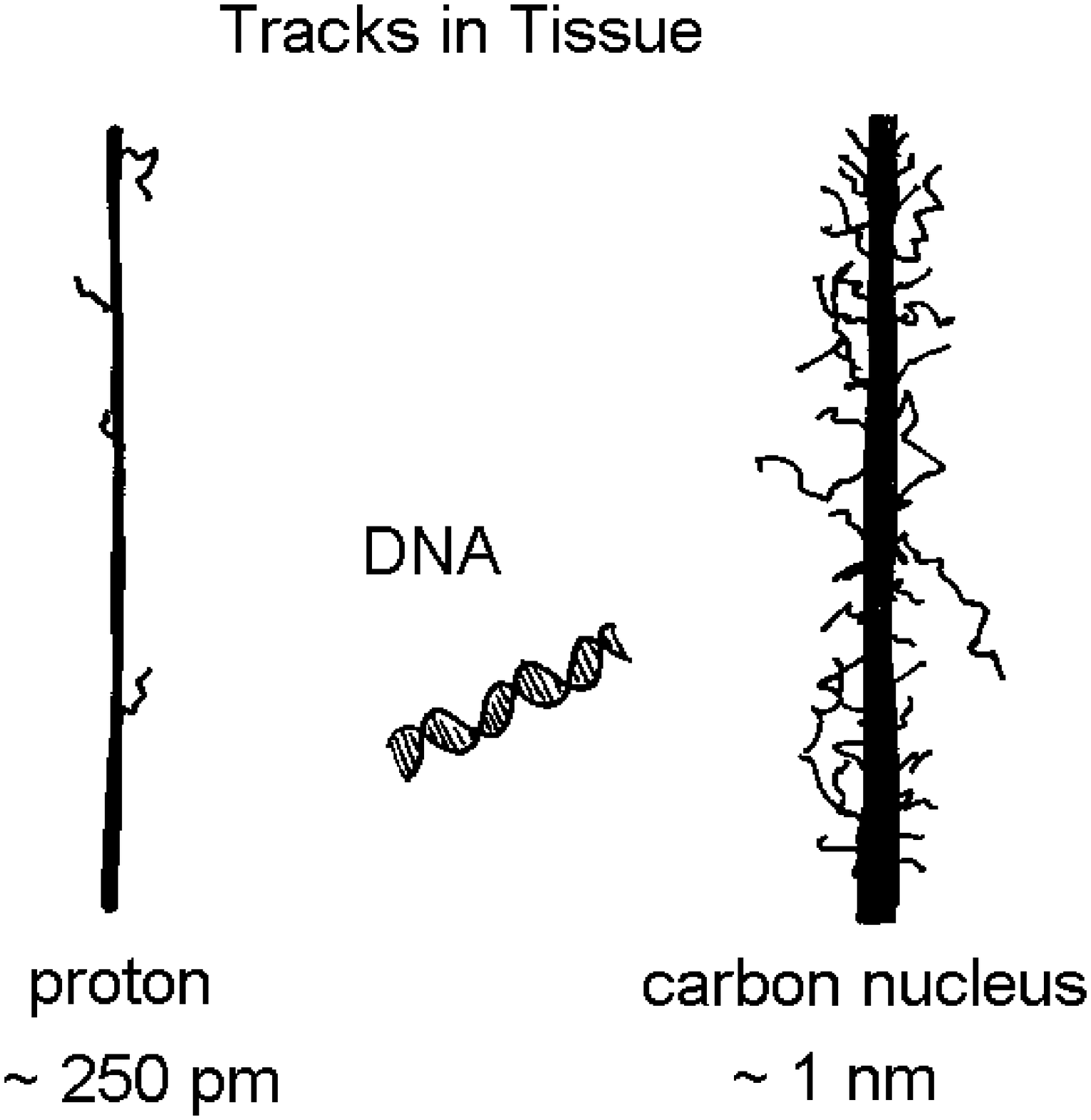} 
\caption{Sketch of a proton and a carbon nucleus track in tissue. The
fuzziness of 
the tracks is caused by short range $\delta$-rays \protect\cite{Kraft1}}
\end{center}
\end{figure}

In addition to ionisation light particles, like electrons, can also undergo 
bremsstrahlung $(dE/dx \sim z^2 Z^2 E)$. Since the probability for this
process is 
inversely proportional to the square of the mass of the beam particle, 
bremsstrahlung can be neglected for particles heavier than the electron for 
energies relevant to tumour therapy \cite{Grupen1}.

The above mentioned fragmentation of heavy ions leads to the production of 
positron emitters.  For the $^{12}$C case, lighter isotopes like $^{11}$C and 
$^{10}$C are produced. Both isotopes decay with short half-lives $(T_{1/2} 
(^{11}C) = 20,38 \; min$; $T_{1/2} (^{10}C) = 19.3 \; s)$ to boron
according to
\begin{eqnarray}
  ^{11}C & \rightarrow & ^{11}B + e^+ + \nu_e    \\
  ^{10}C & \rightarrow & ^{10}B + e^+ + \nu_e \; . \nonumber
\end{eqnarray}
The positrons have a very short range, typically below $1 \; mm$. After
coming to
rest they annihilate with electrons of the tissue giving off two
monochromatic 
photons of $511 \; keV$ which are emitted back-to-back
\begin{equation}
  e^+ + e^- \rightarrow \gamma + \gamma \; .
\end{equation}
These photons can be detected by positron-emission tomography techniques
and can 
be used to monitor the destructive  effect of heavy ions on the tumour tissue.

\section*{Production of particle beams}
 The treatment of deep seated tumours requires charged particles of
typically 100 
 to $400 \; MeV$ per nucleon, i.e. 100 to $400 \; MeV$ protons or 1.2 to
$4.8 \; 
 GeV \; ^{12}$C ions. These particles are accelerated in either a linear 
 accelerator or in a synchrotron. As an example figure~5 shows a typical
set-up 
 for the production of heavy ions. $^{12}$C atoms are evaporated from an ion 
 source and pre-accelerated. Thin foils are used to strip off all electrons
from 
 the ions. The $^{12}$C nuclei are then injected into a synchrotron, where
they 
 are accelerated by radiofrequency cavities to the desired energy. The ions
are 
 kept on track by dipole bending magnets and they are focussed by quadrupoles.
 After having reached the final energy they are ejected by a kicker magnet,
which 
 directs the particles to the treatment room. Their path is monitored by
tracking
 \begin{figure}
\begin{center}
\includegraphics[height=0.4\textheight]{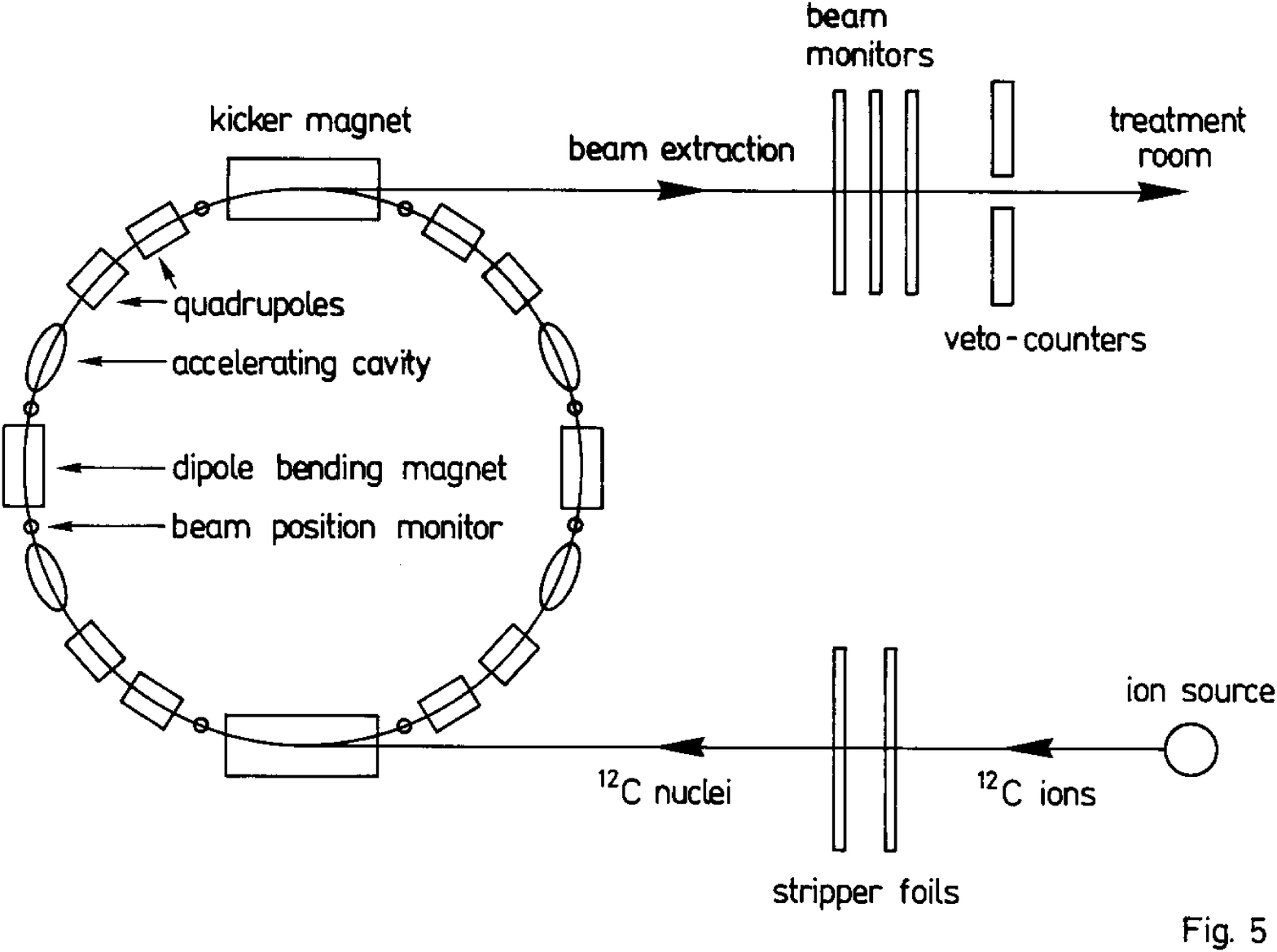}
 \caption{Sketch of a typical set-up for the acceleration of heavy ions
(not all 
 components are shown)}
\end{center}
\end{figure}
chambers (multi-wire proportional counteres, ion chambers or
drift-chambers). If 
beam losses occur veto-counters (mostly scintillation counters) ensure that
only a 
pencil beam is steered to the treatment room.

Nowadays, mainly protons and heavy ions are used for tumour therapy. Other 
possibilities consist of the use of negative pions
\cite{Dyson,Curtis,Goodman},
 which are produced by high energy protons in a beam dump according 
to
\begin{equation}
   p + {\rm nucleus} \rightarrow p + {\rm nucleus} + \pi^- + \pi^+ + \pi^0
\end{equation}
where the $\pi^-$ are momentum selected and collimated. After losing their
energy 
by ionisation the negative pions are 
captured in the tumour tissue by nuclei at the end of their range and produce 
so-called `stars' in which 
neutrons are created. The Bragg-peak of the negative pions along with the
local 
production of neutrons which have a high biological effectiveness leads to an 
efficient  cell killing in the tumour at the end of the pion's range.

Neutrons are also possible candidates for tumour treatment \cite{Lennox}.
For this 
purpose the tumour is sensitized by a boron compound before neutron
treatment. The 
boron compound must be selected in such a way that it is preferentially
deposited 
in the tumour region. Neutrons are then captured by the boron according to:
\begin{equation}
   n + \; ^{10}\! B \rightarrow \; ^7\! Li + \alpha  \; .
 \end{equation}
The produced $\alpha$-particles (He-nuclei) have a very short range ($\sim$ 
several $\mu m$) and high biological effectiveness. Best results are
obtained with 
epithermal neutrons ($\sim 1 \; keV$) produced by $5 \; MeV$ protons  on 
light targets (e.g. Be).

Direct irradiation with neutrons -- without sensitizing the tumour -- has the 
disadvantage that neutrons show a similar dose depth curve like $^{60}$Co 
$\gamma$-rays thus producing a high amount of biologically very effective
damage 
in the healthy tissue around the tumour (see figure 6 \cite{NAC}).

\begin{figure}
\begin{center}
\includegraphics[height=0.3\textheight]{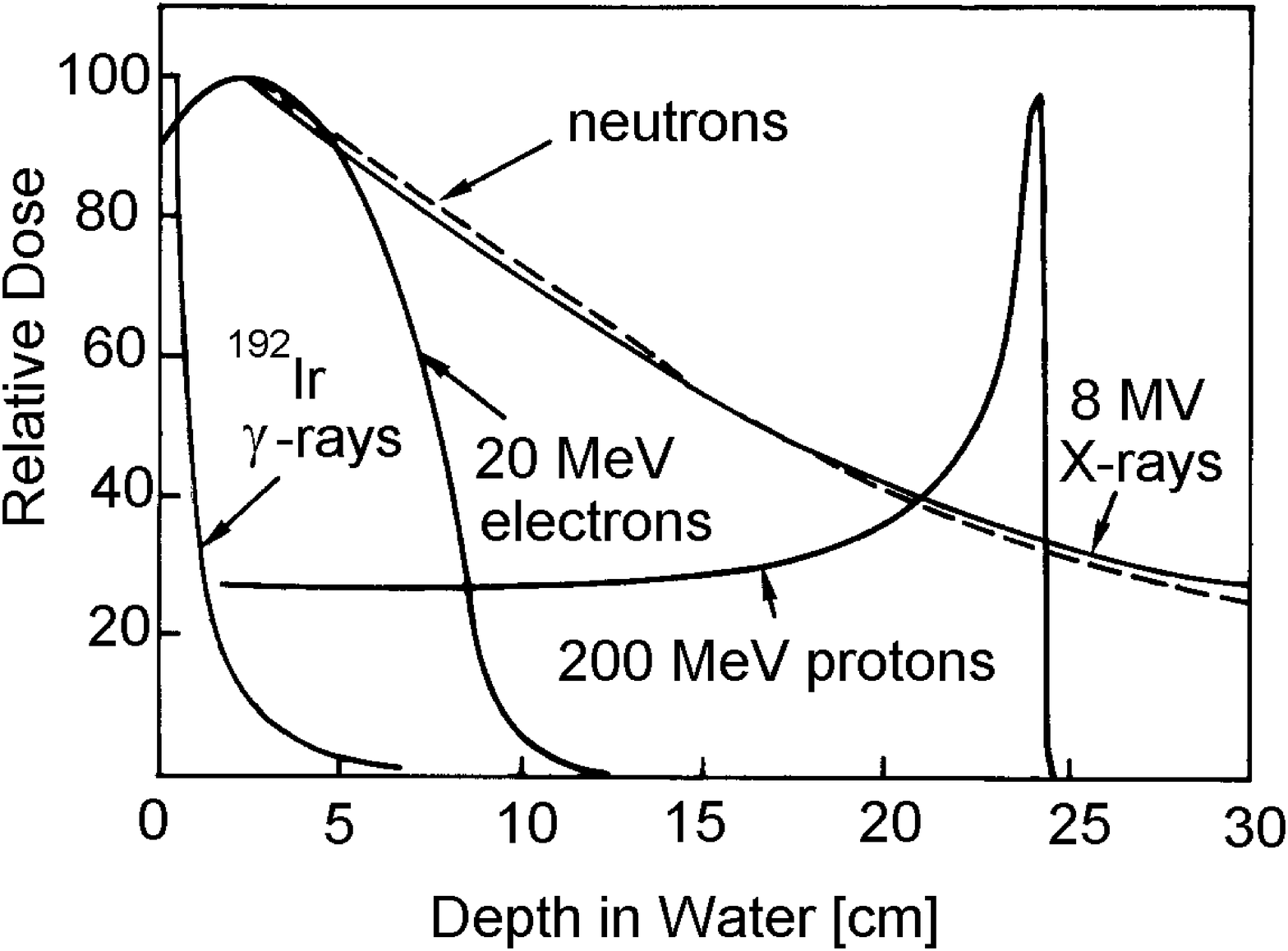}
\caption{Comparison of depth-dose curves of neutrons, $\gamma$-rays
(produced by a
$8 MV$ driven X-ray tube), $200 \; MeV$ protons, $20 \; MeV$ electrons and
$^{192}$Ir-$\gamma$-rays ($161 \; keV$) \protect\cite{NAC}}
\end{center}
\end{figure}

\section*{Applications in Tumour Therapy}
The target for cell killing is the DNA in the cell nucleus (see figure 7
(after 
\cite{Kraft1})). The size of the DNA-molecule compares favorably well with
the 
width of the ionisation track of a heavy ion. The DNA contains two strands
containing identical information. A damage of one strand by ionising
radiation can 
easily be repaired by copying the information from the unaffected strand to
the 
damaged one. Therefore the high ionisation density at the end of a particle's 
range matches well with the requirement to produce double strand breaks in
the DNA, 
which the cell will not survive. Heavy ions like $^{12}$C seem to be
optimal for 
this purpose. Ions heavier than carbon would even be more powerful in
destroying 
tumour tissue, however, their energy loss in the surrounding tissue and in
the 
entrance region already reaches a level where the fraction of irreparable
damage 
is too high, while for lighter ions (like $^{12}$C) mostly repairable
damage is 
produced in the healthy tissue outside the targeted tumour. The cell
killing rate 
in the tumour region thus 
benefits from two properties of protons or ions like carbon:
\begin{figure}
\begin{center}
\includegraphics[height=0.35\textheight]{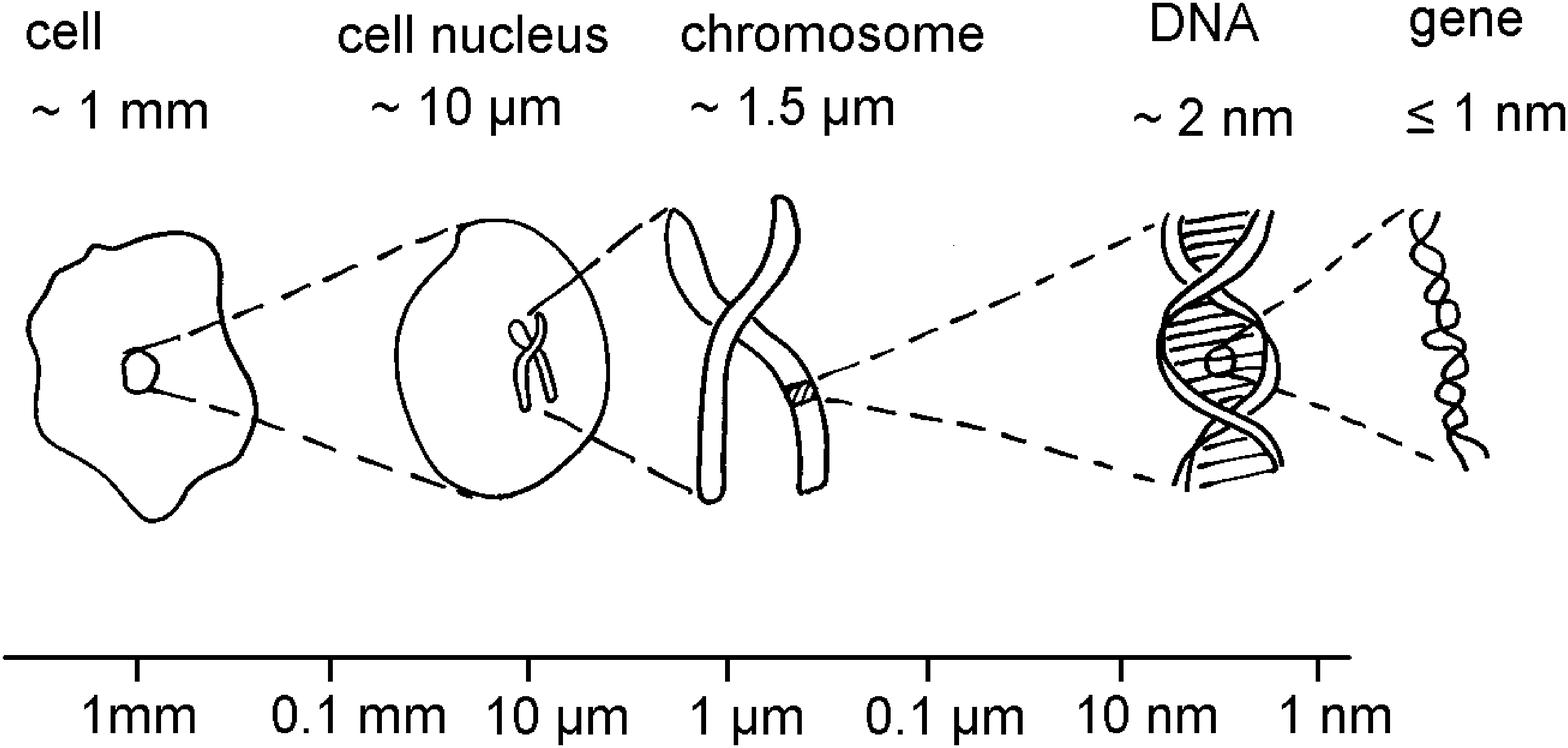}
\caption{Sketch of typical dimensions of biological targets (after 
\protect\cite{Kraft1})}
\end{center}
\end{figure}
\begin{itemize}
\item the increased energy loss of protons and ions at the end of their
range and
\item the increased biological effectiveness of double strand breaks at high 
ionisation density.
\end{itemize}
The cell killing rate is eventually related to the equivalent dose H in the
tumour 
region, which can be expressed by
\begin{equation}
  H = \frac{1}{m} \int \frac{dE}{dx} dx \cdot RBE
\end{equation}
where $m$ is the tumour mass and RBE the increased relative biological 
effectiveness. The integral extends over the tumour region.

As mentioned above the rate and location of cell killing can be monitored 
by observing the annihilation photons which result from the $\beta^+$-decay
of 
fragments formed by the beam.

These physical and biological principles are employed in an efficient way
by the 
raster scan technique \cite{Kraft2,Kraft3,Kraft4}. A pencil beam of 
heavy ions (diameter $\sim 1 \; mm$) is aimed at the tumour. The beam
location and 
spread is monitored by tracking chambers with high spatial resolution. In the 
treatment planning the tumour is subdivided into three-dimensional pixels
(``voxels'').
Then 
the dose required to destroy the tumour, which is proportional to the beam 
intensity, is calculated for every voxel. For a fixed depth in tissue an
areal 
scan is performed by magnetic deflection sweeping the beam across the area
in a 
similar way as a TV image is produced (see figure~8, \cite{Kraft3,Kraft4}). 
The tumour volume is filled from the back by energy variation ($\sim$ range 
variation) of the beam. Typically 50 energy steps are used starting at the
rear 
plane. For a depth profile from $2 \; cm$ to $30 \; cm$ one has to cover
energies 
from $80 \; MeV$/nucleon to $430 \; MeV$/nucleon. When the beam energy is
reduced 
the required dose for the plane under irradiation is calculated using the
damage 
that the more energetic beam had already produced in its entrance region.
This 
ensures that the lateral (caused by magnetic deflection) and longitudinal
scanning
\begin{figure}
\begin{center}
\includegraphics[height=0.275\textheight]{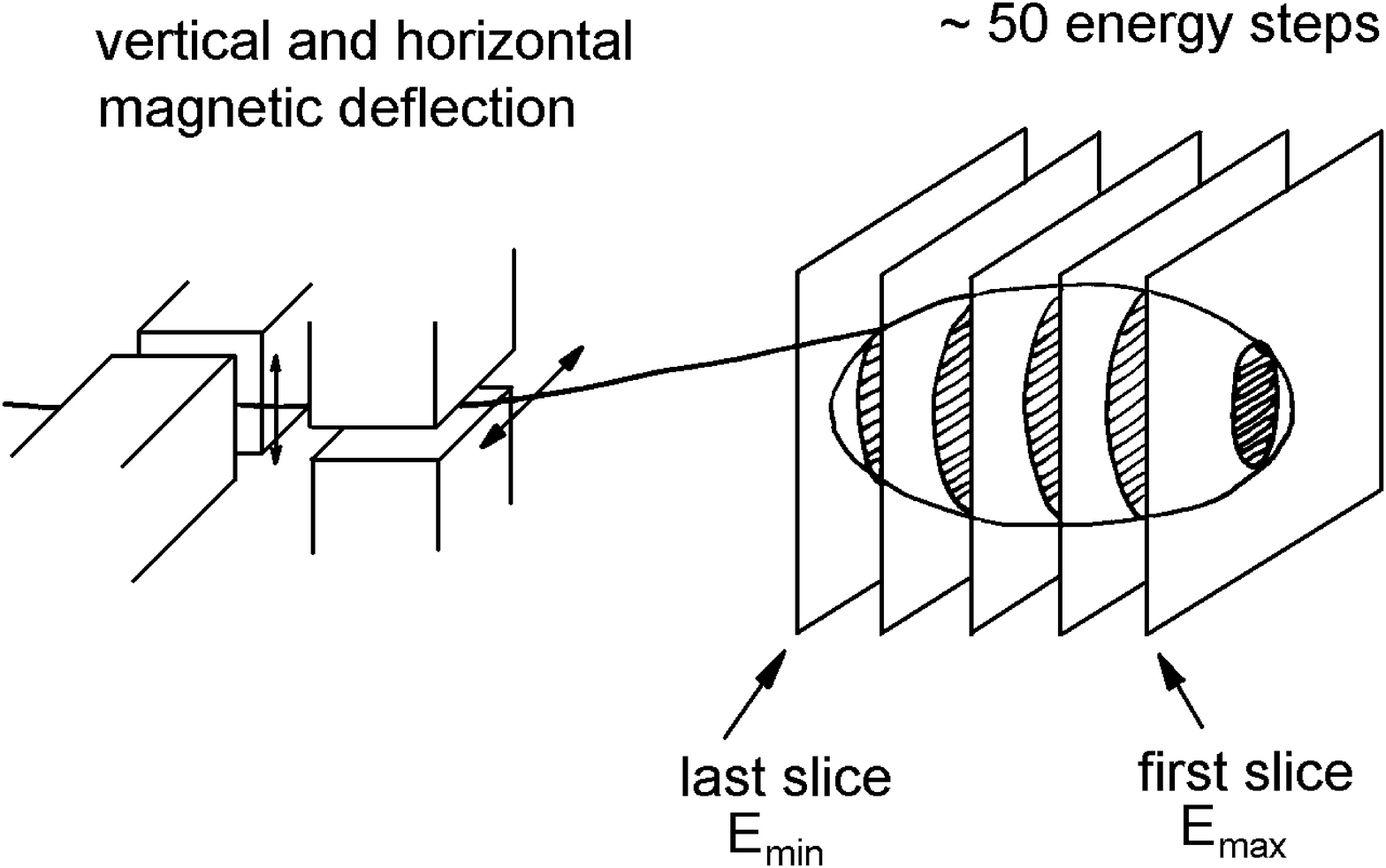}
\caption{Principle of the raster scan method \protect\cite{Kraft3,Kraft4}}
\end{center}
\end{figure}
(by energy variation) covers the tumour completely. In figure 9 (after 
\cite{Kraft1}) the dose distribution for individual energy settings and the 
resulting total dose is sketched and compared with the damage that X-rays
from a
$^{60}$Co-source would produce. An artist impression of the dose
distribution for a 
lung and a brain tumour is given in figure 10.

\section*{Treatment facilities}
Berkeley was the birthplace of therapy with hadrons. Since 1954 protons and
later 
Helium-nuclei were used for treatment. Throughout the world treatment with
protons 
is standard (Sweden, USA, Russia, Japan, Switzerland, England, Belgium,
France, 
South Africa). In some places negative pions have been used in the past (USA, 
Canada, Switzerland). The most promising results have been obtained with
heavy ions
\begin{figure}
\begin{center}
\includegraphics[height=0.35\textheight]{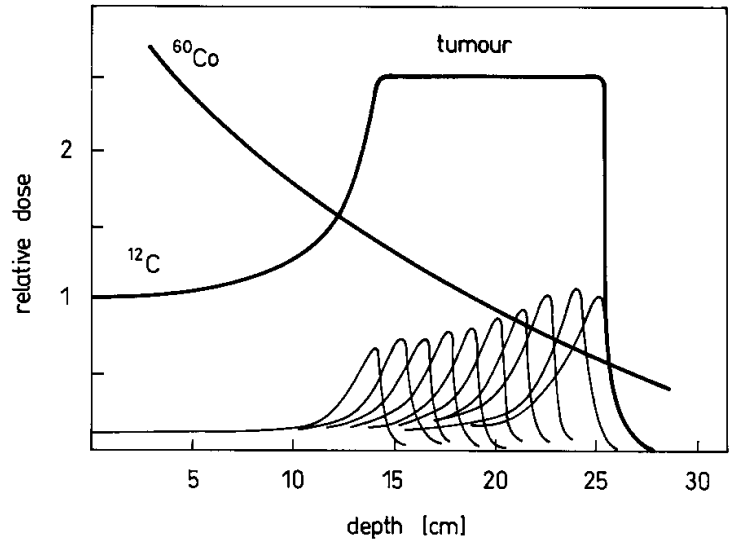}
\caption{Superposition of Bragg-peaks by energy variation (after 
\protect\cite{Kraft1})}
\end{center}
\end{figure}
\begin{figure}
\begin{center}
\includegraphics[height=0.35\textheight]{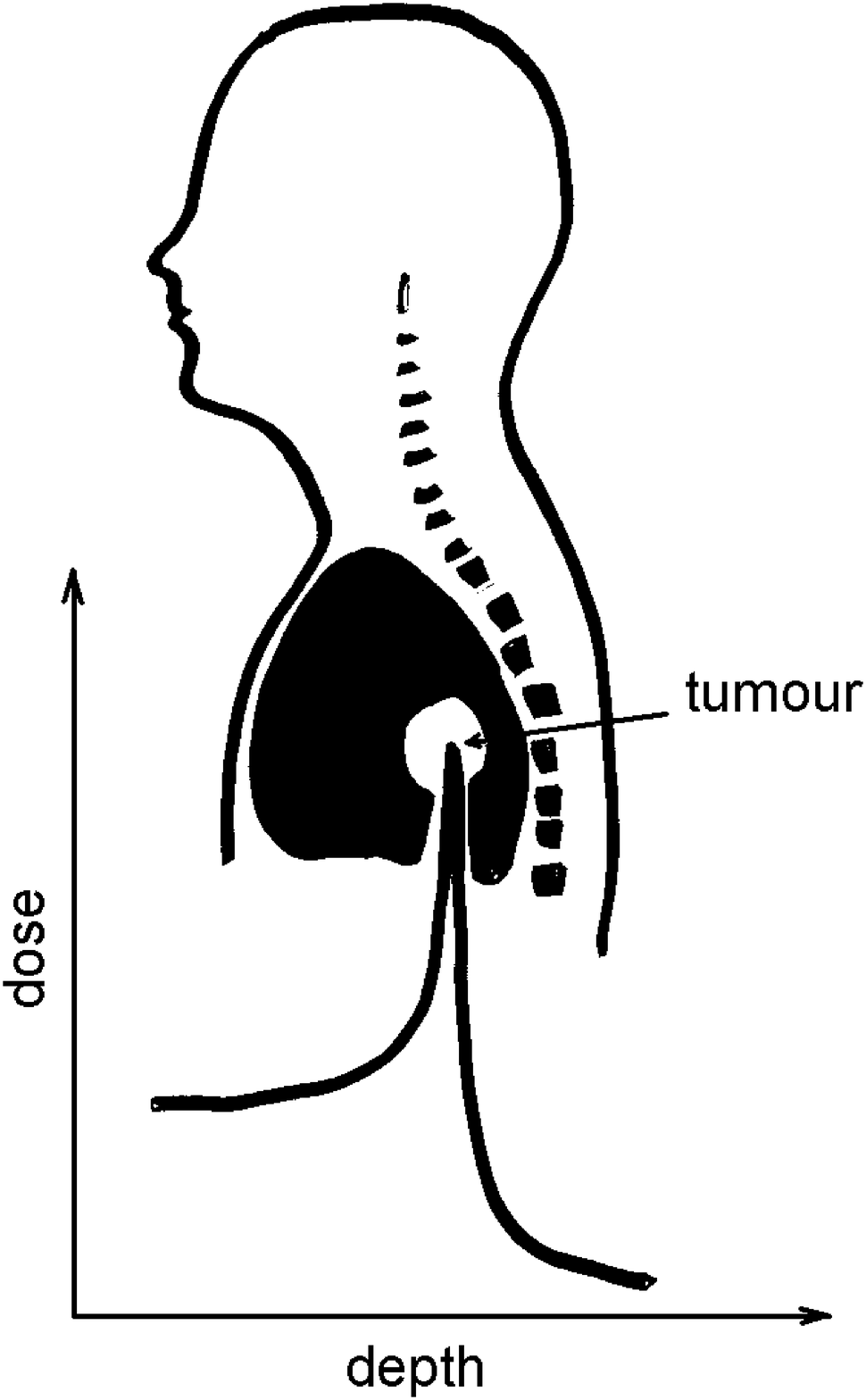}
\hspace{2cm}
\includegraphics[height=0.25\textheight]{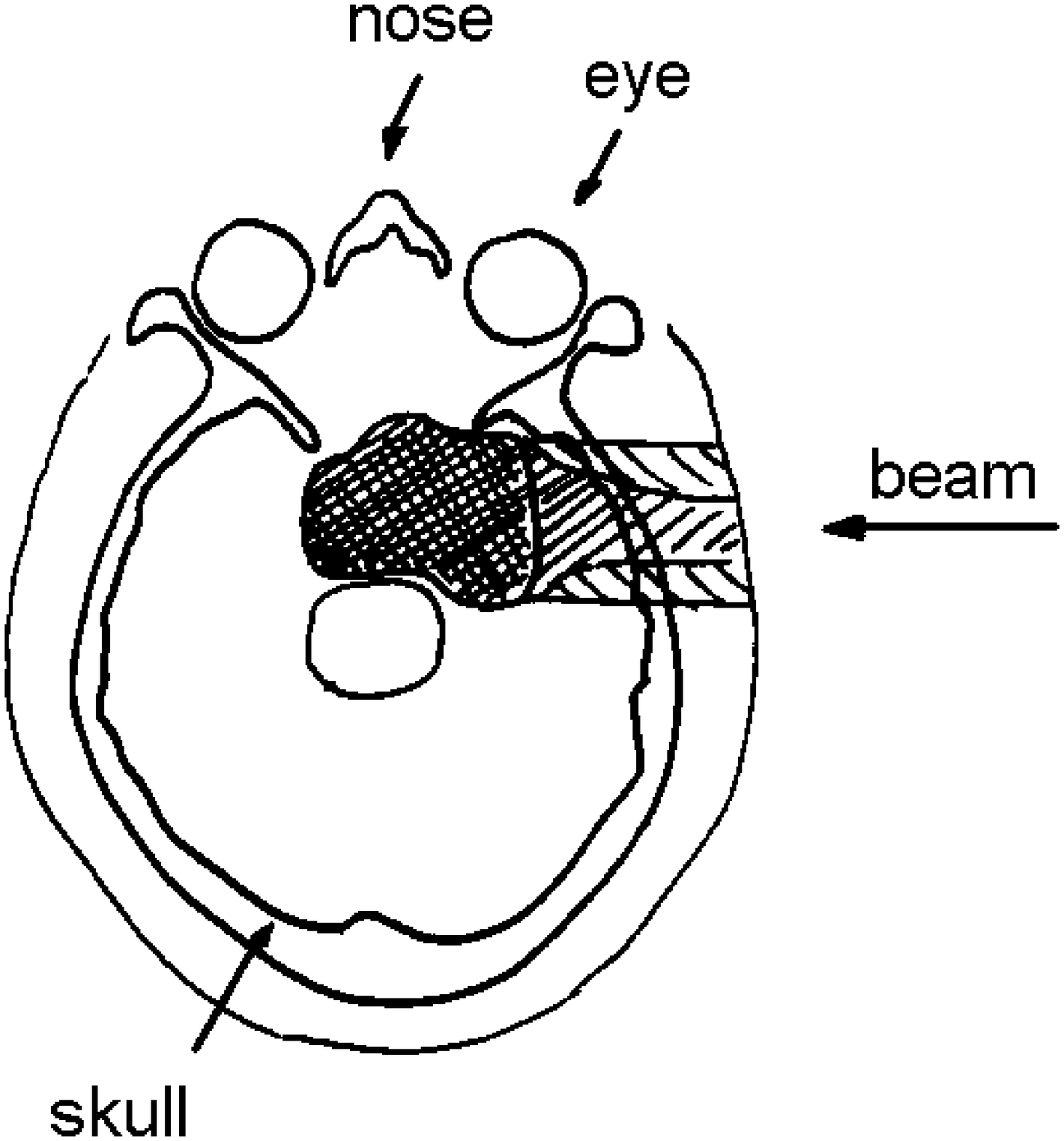}
\caption{a) The position of the Bragg-peak can be adjusted by energy
selection to 
produce a maximum damage at the tumour site (here in the lung).
\protect\linebreak
b) Mapping of a brain tumour with ionisation from heavy ions. Some damage
at the 
entrance region cannot be avoided}
\end{center}
\end{figure}
(Berkeley, USA; Chiba, Japan; and Darmstadt, Germany). In total $\sim$~25000 
patients have been treated from 1954 to 1999.

\section*{Summary and Outlook}
The inverse ionisation dose profile of charged particles has been known for
a long 
time, from nuclear and particle physics. The instrumentation originally
developed for 
elementary particle physics experiments has made it possible to design and
monitor
particle beams with great precision which can then be used for tumour
therapy. 
Heavy ions seem to be ideal projectiles for tumour treatment. They are
suitable 
for well localized tumours. The availability of treatment facilities is 
increasing. Naturally such a facility requires an expensive and complex 
accelerator for the charged particles. For beam steering and control 
sophisticated particle detectors and interlock systems are necessary to
ensure the 
safety of patients.

\section*{Acknowledgements}
The author has benefitted a great deal from information provided by
G.~Kraft from 
GSI-Darmstadt and from discussions with him. I acknowledge also the help of 
Mrs.~L.~Hoppe and C.~Haucke for the drawing of the figures, Mrs.~A.~Wied for 
typing the text, Mr.~Ngac An Bang for giving the paper the final
LaTeX-touch, and
Mr. D. Robinson for a careful reading of the manuscript.

\end{document}